\documentclass[twocolumn,reprint,superscriptaddress,footinbib,aps,pre]{revtex4-2}

\usepackage[utf8]{inputenc}
\usepackage[TS1,T1]{fontenc}

\usepackage{graphicx}

\newcommand{\subfigref}[2]{\ref{#1}\hyperref[#1]{\textbf{(#2)}}}

\usepackage{amsmath,physics}

\usepackage[x11names,table]{xcolor}
\newcommand{\panel}[1]{\textbf{\textcolor{IndianRed3}{#1}}}

\usepackage[]{hyperref}
\hypersetup{
    colorlinks,
    linkcolor={IndianRed3},
    citecolor={Cyan4},
    urlcolor={Cyan4}
}

\def\qd{QD}
\def\resonator{CNT}
\def\NESS{non-equilibrium steady state}
\def\gw{\lambda}
\def\F{F_\text{ex}}
\def\zpm{z_\text{ZPM}}
\def\omegaex{\omega_\text{ex}}
\def\EOM{equation of motion}
\def\P{P_\text{ex}}
\def\I{I_\text{avg}}

\begin{document}

\title{Sources of nonlinearity in the response of a driven nano-electromechanical resonator}

\author{Sofia Sevitz}
\email{sevitz@uni-potsdam.de}
\affiliation{University of Potsdam, Institute of Physics and Astronomy, Karl-Liebknecht-Str. 24-25, 14476 Potsdam, Germany\looseness=-1}

\author{Kushagra Aggarwal}
\affiliation{Department of Materials, University of Oxford, Oxford OX1 3PH, United Kingdom}
\affiliation{Department of Engineering Science, University of Oxford, Oxford OX1 3PJ, United Kingdom}

\author{Jorge Tabanera-Bravo}
\affiliation{Dept.~Estructura de la Materia, F\'isica T\'ermica y Electr\'onica and GISC, Universidad Complutense de Madrid. 28040 Madrid, Spain}
\affiliation{Mathematical bioPhysics group, Max Planck Institute for Multidisciplinary Sciences, Göttingen 37077, Germany}

\author{Juliette Monsel}
\affiliation{Department of Microtechnology and Nanoscience (MC2), Chalmers University of Technology, S-412 96 G\"oteborg, Sweden\looseness=-1}

\author{Florian Vigneau}
\affiliation{Department of Materials, University of Oxford, Oxford OX1 3PH, United Kingdom}

\author{Federico Fedele}
\affiliation{Department of Engineering Science, University of Oxford, Oxford OX1 3PJ, United Kingdom}

\author{Joe Dunlop}
\affiliation{Physics and Astronomy, University of Exeter, Exeter EX4 4QL, United Kingdom}

\author{Juan M.R. Parrondo}
\affiliation{Dept.~Estructura de la Materia, F\'isica T\'ermica y Electr\'onica and GISC, Universidad Complutense de Madrid. 28040 Madrid, Spain}

\author{Gerard J. Milburn}
\affiliation{School of Mathematics and Physics, The University of Queensland, St Lucia QLD 4072, Australia}

\author{Janet Anders}
\affiliation{University of Potsdam, Institute of Physics and Astronomy, Karl-Liebknecht-Str. 24-25, 14476 Potsdam, Germany\looseness=-1}
\affiliation{Physics and Astronomy, University of Exeter, Exeter EX4 4QL, United Kingdom}

\author{Natalia Ares}
\affiliation{Department of Engineering Science, University of Oxford, Oxford OX1 3PJ, United Kingdom}

\author{Federico Cerisola}
\email{f.cerisola@exeter.ac.uk}
\affiliation{Physics and Astronomy, University of Exeter, Exeter EX4 4QL, United Kingdom}

\date{\today} 

\begin{abstract}

Nanoelectromechanical resonators provide an ideal platform for investigating the interplay between electron transport and nonlinear mechanical motion. Externally driven suspended carbon nanotubes, containing an electrostatically defined quantum dot are especially promising. These devices possess two main sources of nonlinearity: the electromechanical coupling and the intrinsic contributions of the resonator that induce a Duffing-like nonlinear behavior. In this work, we observe the interplay between the two sources across different driving regimes. The main nonlinear feature we observe is the emergence of arch-like resonances in the electronic transport when the resonator is strongly driven. We show that our model is in good agreement with our experimental electron transport measurements on a suspended carbon nanotube. This characterization paves the way for the exploration of nonlinear phenomena using mesoscopic electromechanical resonators.

\end{abstract}

\maketitle

\section{Introduction}\label{Sec:Introduction}

Nanomechanical resonators have emerged as a unique platform for the exploration of nonlinear dynamics. They offer insights into various nonlinear phenomena, such as spontaneous self-oscillations~\cite{Eichler2011, Wen2019, Urgell2019} and bistable states of mechanical motion~\cite{Tabanera2024, Belardinelli2023}. For this reason, they have been suggested for wide range of applications across different scientific fields, ranging from thermodynamics~\cite{Tesser2022} to chaos~\cite{Gusso2019} and metrology~\cite{Woolley2008}. 

\begin{figure}[t]
    \centering
    \includegraphics[width=\linewidth]{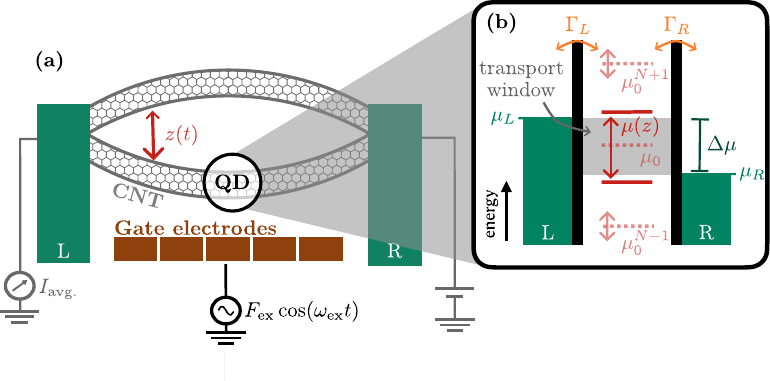}
    \caption{
    \textbf{(a)} Schematic of the platform: a carbon nanotube (\resonator) is suspended between the left (L) and right (R) metal electrodes that act as fermionic reservoirs for the embedded quantum dot (\qd). The \qd~is electrostatically defined via the gate electrodes. Furthermore, in one of the gates, an external AC drive with force $\F$ and frequency $\omega_\text{ex}$ excites the displacement $z$ of the \resonator. Application of a bias voltage $\Delta\mu$ between the L/R reservoir drives a current $\I$ through the device. 
    \textbf{(b)} Sketch of the electrochemical potential structure of the embedded QD characterized by  electrochemical potentials $\mu_0^N$. Due to the electromechanical coupling, the electrochemical potential acquires a displacement dependence $\mu(z)$. We include two additional electrochemical potentials of the QD, $\mu_0^{N-1}$ and $\mu_0^{N+1}$, that become relevant when comparing with experimental results in Sec.~\ref{Sec:Multilevel_gen}. The L and R fermionic reservoirs have electrochemical potentials $\mu_L$ and $\mu_R$, respectively, which enable a transport window where electrons in the \qd~can \textit{hop} from L to R or vice versa with tunneling rates $\Gamma_L$ and $\Gamma_R$.
    }
    \label{fig:Figure1}
\end{figure}

One type of resonators of particular interest are suspended carbon nanotubes (\resonator) due to their large zero-point motion~\cite{Chaste2012, Wang2017} and large quality factors~\cite{Moser2014}. Here we study a suspended \resonator~device that is externally driven, see Fig.~\ref{fig:Figure1}. It is possible to electrostatically define a quantum dot (\qd) along the \resonator, and its electronic occupation is strongly affected by the motion of the \resonator~\cite{Zhong2008, Park2004, Zhou2005, Laird2015}. In turn, the motion of the \resonator~is altered by the electronic tunneling through the \qd. Indeed, it has recently been reported that the coupling between electron tunneling and mechanical motion is in the ultrastrong regime, resulting in strong back action~\cite{Vigneau2022, Samanta2023}.

In the platform studied here, there are two distinct sources of nonlinearities affecting the mechanical motion: (i) the back-action due to the electromechanical coupling and (ii) intrinsic mechanical nonlinearities. The contributions of each source depends strongly on the driving power and on other device operation parameters such as gate voltages~\cite{Steele2009,Samanta2018,Steeneken2021}. Typically, at low driving power, source (i) dominates~\cite{Steele2009, Meerwaldt2012, Meerwaldt2012Chapter}. Meanwhile, source (ii) starts to play a more predominant role at higher driving powers. Signatures of this second nonlinearity have been previously observed experimentally presenting as a Duffing-type response~\cite{Steele2009, Wang2021, Willick2017}. Although a lot of efforts have been made to model each nonlinear source independently~\cite{Wang2021,Samanta2023,Luo2017,Cong2021,Kirton2013,Meerwaldt2012,Steele2009,Witkamp2006,Sapmaz2003,Ho1975,Conley2008,Westra2010,Eichler2012,Kozinsky2006}, to our knowledge, the interplay between these two effects have not been yet explained.

In this work, we capture these nonlinear contributions within the same microscopic model for a single high-frequency mechanical mode of a carbon nanotube resonator. For this, we develop a theoretical model from first principles valid across different external driving regimes. We are specifically interested in the response of the system in the \NESS. Here we fully characterize the contributions of the nonlinear sources and observe their features in the amplitude of the \resonator~and the time-averaged electronic current through the device. First we study the case where a single transport channel plays a role in the electronic transport. When the \resonator~is weakly driven, source (i) is predominant and we recover the typical frequency softening due to the \qd--\resonator~coupling~\cite{Steele2009, Meerwaldt2012}. As the driving of the \resonator~increases and source (ii) starts to become dominant, arch-like resonance structures appear in the electronic current. This appearance is the main feature of our model. To compare our theoretical findings with the experimental results presented here, we generalize our model to account for multiple transport channels. We find that our theoretical model agrees well with experimental results. This characterization is essential to pave the way for the exploration of fundamental experiments, such as observing signatures of quantum to classical transition at the nanoscale~\cite{Katz2007}.

\section{Single electrochemical potential model}\label{Sec:Model}

\subsection{Microscopic model}

The system is composed of an externally driven suspended \resonator~which is free to vibrate with displacement $z$, see Fig.~\subfigref{fig:Figure1}{a}. A \qd~is electrostatically defined along the \resonator. For now, we consider this \qd~to have a single transport channel, i.e an electrochemical potential $\mu_0$ within the bias window. The \qd~is coupled to the mechanical motion~\cite{Steele2009, Meerwaldt2012, Vigneau2022, Meerwaldt2012Chapter}. In Sec.~\ref{Sec:Multilevel_gen}, we will generalize the single electrochemical potential model to the multiple level case, as depicted in Fig.~\subfigref{fig:Figure1}{b}. The full system is described by the spinless Anderson-Holstein model, with additional mechanical terms to model the intrinsic nonlinearities of the resonator and the external driving. We have
\begin{align}\label{eq:full_Hamiltonian}
    \hat{H} &= \hat{H}_\text{AH} + \hat{H}_\text{duff}+\hat{H}_\text{drive},
\end{align}
where first term constitutes the standard Anderson-Holstein model~\cite{Albrecht2013,Wang2021,Mitra2004, Piovano2011}
\begin{equation}\label{eq:H_AH}
    \hat{H}_\text{AH}= \hat{H}_\text{\qd}  + \hat{H}_\text{leads} +\hat{H}_\text{T} + \hat{H}_\text{\resonator} + \hat{H}_\text{\resonator-\qd}.
\end{equation}
Here, the first term is the Hamiltonian of the \qd, $\hat{H}_\text{\qd}= \mu_0 \hat{n}$, with occupation $\hat{n}= \hat{d}^\dagger \hat{d}$ where $\hat{d}^\dagger$($\hat{d}$) is the creation (annihilation) operator of the dot. Electronic reservoirs are located to the left ($L$) and right ($R$) of the dot and are governed by the Hamiltonian $\hat{H}_\text{leads}=\sum_{\nu} \epsilon_\nu \hat{c}_\nu^\dagger \hat{c}_\nu$ where $\epsilon_\nu$ are the energies of the electrons in reservoir $\nu$ and the operators $\hat{c}_\nu^\dagger$($\hat{c}_\nu$) are their respective creation (annihilation) operators. Throughout this work we consider these reservoirs to always be in local thermal equilibrium. Eq.~\eqref{eq:H_AH} further models single-electron transport between the dot and the reservoirs by the tunneling Hamiltonian $H_\text{T}=\sum_\nu t_{\nu} \hat{c}_\nu^\dagger \hat{d} + \text{h.c}.$, where $t_{\nu}$ is the coupling constant related to the tunneling rate $\Gamma_\nu$. We take energy-independent rates also known as the \textit{wide-band limit}~\cite{Moulhim2021, Potts2024, Schaller2014Book}. The third term of Eq.~\eqref{eq:H_AH} corresponds to the linear dynamics of the \resonator~motion, $\hat{H}_\text{\resonator}=\frac{1}{2m} \hat{p}^2 + \frac{1}{2} m\omega^2 \hat{z}^2$, where $\hat{p}$ and $\hat{z}$ are the momentum and displacement operators of the mechanics, respectively, $m$ is the mass, and $\omega$ its natural frequency. Finally, the last term of Eq.~\eqref{eq:H_AH} is the linearised coupling between the mechanics and the electronic transport, given by~\cite{Vigneau2022,Mitra2004, Piovano2011}
\begin{equation}\label{eq:H_CNT_QD}
    \hat{H}_\text{\resonator--\qd}= \hbar \omega \gw \frac{\hat{z}}{z_\text{ZPM}} \hat{n},
\end{equation}
where the mechanical zero-point motion is $z_\text{ZPM}= \sqrt{\hbar/2m\omega}$ and the unit-free parameter $\gw$ quantifies strength of this electromechanical coupling. If $\gw \ll 1$, the \qd~and the \resonator~motion are weakly coupled and thus can be treated perturbatively. On the contrary, if $\gw \gtrsim 1$, we are in the ultrastrong coupling regime, where the electromechanical interaction significantly affects both the \qd~and the \resonator~motion. In what follows, we will focus on the ultrastrong case, in agreement with recent experimental characterizations of these devices~\cite{Vigneau2022, Samanta2023}. Due to the coupling between the mechanics and the dot, the electrochemical potential of the dot depends on the displacement $z$. We take here a first-order expansion in displacement as~\cite{Vigneau2022}, 
\begin{equation}\label{eq:Chemical_eff}
     \mu(z) = \mu_0 + \hbar \omega \gw  \frac{\hat{z}}{z_\text{ZPM}}.
\end{equation}

An additional term in the total Hamiltonian, see Eq.~\eqref{eq:full_Hamiltonian}, is included to go beyond the standard Anderson-Holstein model and capture the intrinsic mechanical non-linearity of the \resonator, which we here take as a Duffing term~\cite{Meerwaldt2012Chapter,Steeneken2021}
\begin{equation}\label{eq:H_duff}
    \hat{H}_\text{duff}=\frac{\beta}{4}  \left(\frac{\hat{z}}{z_\text{ZPM}}\right)^4,
\end{equation}
where $\beta$ quantifies the strength of these nonlinearities. 

Finally, the last term of Eq.~\eqref{eq:full_Hamiltonian} describes the external AC drive with force $\F$ and frequency $\omegaex$,
\begin{equation}\label{eq:H_drive}
    \hat{H}_\text{drive}= \F \cos({\omegaex t}) \hat{z}.
\end{equation}

This model captures the two physical nonlinearities of the system: (i) due to the electromechanical coupling [Eq.~\eqref{eq:H_CNT_QD}] and (ii) from the intrinsic mechanical nonlinearities [Eq.~\eqref{eq:H_duff}].

For the reminder of this article, we will focus on the \NESS~of the system. For this, we consider the \textit{quasi-adiabatic limit} where the oscillations of the resonator are slow compared to the electron tunneling, i.e. $\omega \ll \Gamma_L, \Gamma_R $. In this regime, a semi-classical model can be employed~\cite{Mitra2004,Clerk2005,Vigneau2022} which is given as follows. First, we neglect the higher order moments of the oscillator and consider only the average position $\langle \hat{z} \rangle = z(t)$. Thus, from the full Hamiltonian detailed in Eq~\eqref{eq:full_Hamiltonian}, the displacement with respect to the zero-point motion $x(t)=z(t)/\zpm$ obeys a classical equation of motion
\begin{equation}\label{eq:displacement_EOM} 
     \ddot{x} + \delta \dot{x} + \omega^2 \Big[x+2\gw n(x) \Big] + \epsilon x^3 = f_\text{ex} \cos(\omegaex t),
\end{equation}
where $\epsilon=\beta/(m \zpm^2)$ is the normalized Duffing parameter and  $f_\text{ex}=\F/(m \zpm)$ is the normalized external driving;
see Appendix~\ref{Appendix:EOM_response} for a detailed derivation. Here, we also have included a damping term $\delta\dot{x}$ to properly account for the intrinsic damping of the \resonator~motion. Note that in Eq.~\eqref{eq:displacement_EOM} we retain the combined effect of both the intrinsic and transport-induced nonlinearities, which as we will see are necessary to properly capture observed behavior of the \resonator~when strongly driven. In Eq.~\eqref{eq:displacement_EOM} $n$ is the average population of the dot, i.e. $\langle \hat{n}\rangle = n(t)$. We will assume that displacement-dot correlations can be neglected, that is $\langle \hat{z}\hat{n}\rangle = \langle\hat{z}\rangle \langle\hat{n}\rangle$, similarly to previous works~\cite{Vigneau2022}. Typical \resonator~experiments operate in the \textit{lifetime broadened regime}, where the electronic tunneling rates $\Gamma_\nu$ are much larger than the fermionic reservoir temperatures $k_B T_\nu$. In such case, \qd~rate equations are not strictly valid and a more sophisticated approach is needed. We follow Refs. \cite{Moulhim2021,Schaller2014Book}, where it is shown that under the \textit{wide-band limit} and $T_L,T_R\simeq 0$ assumptions, the average population of the dot is given by
\begin{equation}\label{eq:population_eq}
    n(x)= \frac{1}{2}+ \frac{\Gamma_R}{\pi \Gamma_\text{tot}}\Big[\arctan\Big(\gamma_R(x) \Big)+\frac{\Gamma_L}{\Gamma_R}\arctan\Big(\gamma_L(x)\Big)\Big],
\end{equation}
with $\gamma_\nu(x)=2(\mu_\nu-\mu(x))/(\hbar \Gamma_\text{tot})$ and $\Gamma_\text{tot} = \Gamma_L+\Gamma_R$ is the total tunneling rate.
We note that due to the explicit dependence of the \qd~electrochemical potential on $x$, an implicit time dependence is also acquired via $x(t)$. 

We are now ready to characterize the \NESS~response of the \qd~and \resonator~motion. In the following section, Sec.~\ref{Sec:NESS_response}, we will analyze the displacement of the \resonator~for different driving regimes. Then, in Sec.~\ref{Sec:el_current}, we study the response of the \qd~population by computing the electron current flowing throught the device.

\subsection{Non-equilibrium steady state response} \label{Sec:NESS_response}

\begin{figure}[t]
    \centering
    \includegraphics[width=\linewidth]{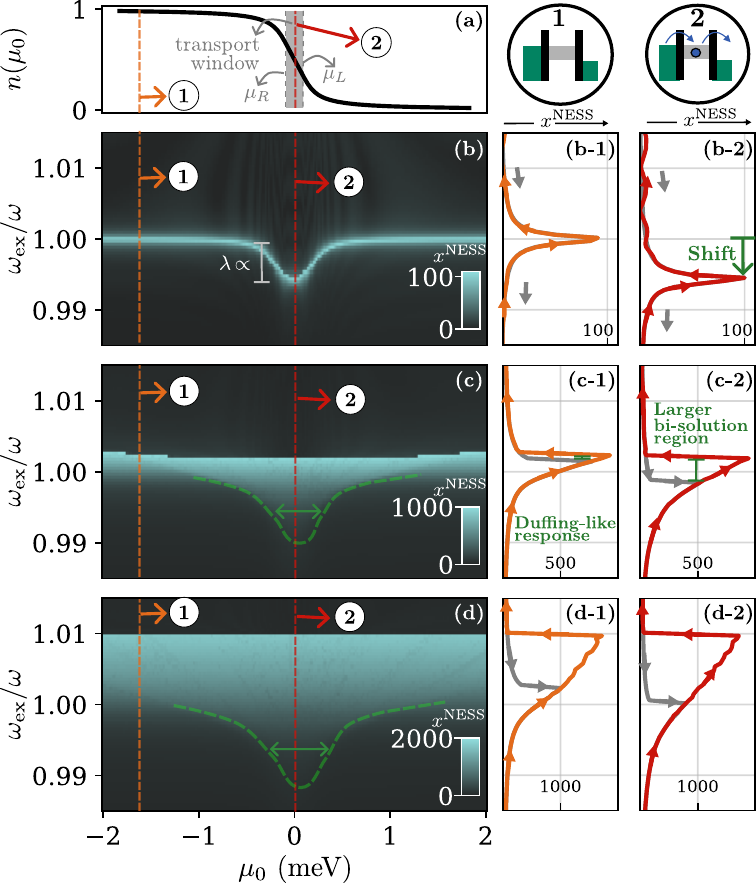}
    \caption{
    \textbf{(a)} \qd~population as a function of the charging energy $\mu_0$. The shaded gray region identifies the values of $\mu_0$ where electronic transport is enabled. For simplicity, the \resonator~motion's back action is not taken into account, i.e. $\gw =0 $.  
    \textbf{(b)-(d)} Displacement of the \resonator~motion with respect to its zero point motion $x^\text{NESS}=z^\text{NESS}/\zpm$ in the \NESS~as a function of both the external driving frequency $\omegaex$ and the bare \qd~energy $\mu_0$ for increasing driving forces as: \textbf{(b)} $\F=0.4$ fN, \textbf{(c)} $\F=4$ fN and \textbf{(d)} $\F=8$ fN. In these panels the driving frequency $\omegaex$ swept from low to high values. 
    \textbf{(b-1)}-\textbf{(c-1)} Frequency response of the \resonator~for a constant \qd~energy of $\mu_0=-1.6$ meV (where no transport is present) highlighted as \textbf{(1)} in panels \textbf{(b)}-\textbf{(d)}. In orange the $\omegaex$ is swept from low to high values while the opposite sweeping direction is presented in gray. 
    Similarly panels \textbf{(b-2)}-\textbf{(c-2)} also showcase the frequency response of the \resonator~but for $\mu_0=0$ meV  (where transport is enabled) highlighted with \textbf{(2)} in panels \textbf{(b)}-\textbf{(d)}. In this case $\omegaex$ is swept from low to high values in red while the opposite sweeping direction is again presented in gray.
    Parameters: $\zpm=1$ pm , $\omega=100$ MHz, 
    $m=0.5$ fg, $\delta=70 $ kHz, $\beta=4.69 \times 10^{-5}$ qJ, $\gw=$ 1.73, $\mu_L=-\mu_R= 0.1$ meV and $\Gamma_L/2\pi=\Gamma_R/2\pi= 30$ GHz.
    }
    \label{fig:Figure2}
\end{figure}

In this section, we study the \NESS~displacement of the \resonator. For this, we simultaneously solve Eq.~\eqref{eq:displacement_EOM}-\eqref{eq:population_eq}, see Appendix~\ref{Appendix:Num_details} for details. 

In Fig.~\subfigref{fig:Figure2}{a} we plot the bare population of the \qd, i.e. setting $\gw=0$, to characterize the regions of $\mu_0$ where electronic transport is enabled (highlighted in gray). Then, taking the electronic back action into account, i.e. $\gw\neq0$, we turn our attention to the non-equilibrium steady-state displacement of the \resonator, $x^\text{NESS}$. This is plotted in Fig.~\subfigref{fig:Figure2}{b}-\subfigref{fig:Figure2}{d}, as a function of the driving frequency $\omega_\text{ex}$ and the bare \qd~energy $\mu_0$ for different driving forces.

Fig.~\subfigref{fig:Figure2}{b} corresponds to the weak driving regime where $\omegaex$ is swept from low to high values. If we trace a line cut in a region where $\mu_0$ is outside the transport window, labeled as \textbf{(1)}, we observe that the \resonator~responds as a driven harmonic oscillator (see panel \panel{(b-1)}). As $\mu_0$ approaches the transport window, situation labeled as \textbf{(2)}, the resonance frequency presents a softening dip shifting the resonance peak to lower $\omegaex$ (see panel \panel{(b-2)}). The height of this dip depends on the coupling ratio $\gw$. At higher $\gw$, nonlinearities arise due to the transport back action, see also Ref.~\cite{Meerwaldt2012Chapter}.

As the external driving force is increased, Fig.~\subfigref{fig:Figure2}{c}, the \resonator~maximum displacement increases and the frequency response broadens. At an intermediate driving force, both sources of nonlinearities play a role. Inside the transport regime, the softening dip is still present but wider (highlighted in green). If we again make a line cut for values of $\mu_0$ outside \panel{(c-1)} and inside \panel{(c-2)} the transport window, we note that the frequency response resembles a typical Duffing oscillator with a sharp drop for $\omegaex > \omega$~\footnote{This has to do with the sign chosen for $\beta$, a negative $\beta$ would result in a sharp drop for $\omegaex<\omega$ (see Supplementary Material).}. Furthermore, hysteresis is present in both situations: the response is different as a function of the external frequency direction sweep. This is a consequence of the bistability caused by the non-linear Duffing term. Note that the region of bistability is increased when electronic transport is enabled compared to when it is not. This is an important observation as it is a key feature for some applications such as to observe quantum-to-classical transitions~\cite{Katz2007}. 

Finally, in Fig.~\subfigref{fig:Figure2}{d} we present the strongly driven regime. Here, the intrinsic nonlinearities of the \resonator~motion dominate, thus the Duffing-like response is accentuated, though the broadened frequency dip due to the electronic back action is still present (highlighted in green). 

Now that we have a good understanding of the different responses of the \resonator~motion for different driving regimes, we move on to study in the next section the signatures of the \qd~response in the electronic current flowing through the device which is experimentally accessible.

\begin{figure}[t]
    \centering
    \includegraphics[width=\linewidth]{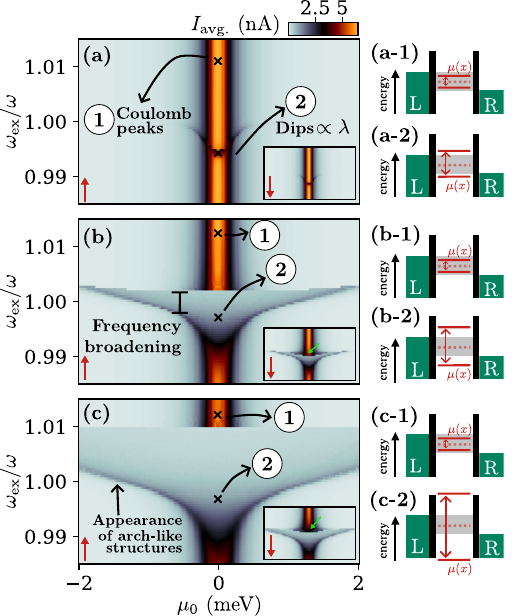}
    \caption{Average electronic current $\I$ for three increasing forces labeled as \textbf{(a)} $\F=0.4$ fN, \textbf{(b)} $\F=4$ fN and \textbf{(c)} $\F=8$ fN as a function of the driving frequency $\omegaex$ and \qd~bare electrochemical potential $\mu_0$. In the main panel $\omegaex$ is swept from low to high values while in the inset the $\omegaex$ is swept in the opposite direction as indicated with red arrows. 
    \textbf{(a-1)-(c-1)} Energy sketch of the \qd~when the $\omegaex$ is far from the \resonator~bare frequency $\omega$.
    \textbf{(a-2)-(c-2)} Same as \textbf{(a-1)-(c-1)} but for $\omegaex$ close to $\omega$ and thus $\mu$ is strongly effected by the displacement of the \resonator.
    Parameters: $\zpm=1$ pm , $\omega=100$ MHz, 
    $m=0.5$ fg, $\delta=70 $ kHz, $\beta=4.69 \times 10^{-5}$ qJ, $\gw=$ 1.73, $\mu_L=-\mu_R= 0.1$ meV and $\Gamma_L/2\pi=\Gamma_R/2\pi= 30$ GHz.
    }
    \label{fig:Figure3}
\end{figure}

\subsection{Measurable observable: the electronic current}\label{Sec:el_current}

In this section we focus on the \NESS~response of the \qd. Instead of analyzing the dot population [Eq.~\eqref{eq:population_eq}], we compute the electronic transport through the device which allows us to easily compare with experiments. Again, following Refs.~\cite{Schaller2014Book, Moulhim2021}, the instantaneous current though the device is
\begin{equation}\label{eq:current_eq}
    I (x) =  \frac{e}{\pi} \frac{\Gamma_L\Gamma_R}{\Gamma_\text{tot}} \Big[\arctan\left(\gamma_L(x)\right)-\arctan\left(\gamma_R(x)\right)\Big].
\end{equation}
where the parameters are the same as presented in Eq.~\eqref{eq:population_eq}. In experiments, what is measured is the averaged current over various oscillation periods in the \NESS. Therefore, for the reminder of the work we will focus on the average current
\begin{equation}\label{eq:Average_current}
    \I= \lim_{\tau \rightarrow \infty} \int_\tau^{\tau+4\pi / \omega} \mathrm{d}t\, I(x(t)),
\end{equation}
with $x(t)$ given by the non-equilibrium steady-state solution of Eq.~\eqref{eq:displacement_EOM}. In Fig.~\ref{fig:Figure3} we plot the average current as a function of the driving frequency $\omega_\text{ex}$ and the bare \qd~energy $\mu_0$ for different driving forces. In the main plots $\omega_\text{ex}$ is swept from low to high values while the insets show the opposite case. In the weak driving regime, corresponding to panel \panel{(a)}, away from the mechanical resonance, Coulomb peaks are observed~\cite{Beenakker1991}. When the driving frequency approaches the natural frequency of the resonator, $\omegaex\simeq\omega$, a softening dip in the current is present, as previously observed in multiple experiments~\cite{Steele2009, Vigneau2022, Meerwaldt2012, Lassagne2009}. The behavior is in accordance with the resonance frequency softening dip of the \resonator~motion, see Fig.~\subfigref{fig:Figure2}{b}. The reason for this decrease of $\I$ is due to the fact that the back action of the \resonator~motion displaces the electrochemical potential of the \qd~outside of the transport window (see Fig.~\subfigref{fig:Figure3}{a-2}), and thus in average the current is reduced.

Next, we increase the driving force, see Fig.~\subfigref{fig:Figure3}{b}. We reproduce experimental observations made in Refs.~\cite{Steele2009,Willick2017} where the frequency dip is broadened with a sharp current change around $\omegaex/\omega\simeq 1.05$. Similar to the displacement plot (Fig.~\subfigref{fig:Figure2}{c}), the contribution of both sources of nonlinearities play a role. For this driving regime, the \resonator~displacement presented a hysteresis behavior, which is also observed in $\I$. This is clearly seen by comparing the main panels of Fig.~\ref{fig:Figure3} to the insets, were the $\omegaex$ is swept in the opposite direction. Indeed, when $\omegaex$ is swept from high to low values, instead of a sharp stop at $\omegaex /\omega \simeq 1.05$ a small dip (marked by a green arrow) is present, which persists for higher driving forces. 

Finally, in the strong driving regime, Fig.~\subfigref{fig:Figure3}{c}, the transport response presents a very distinct feature. The broadening of the frequency outside the transport window is distorted as an arch-like shape, where inside this resonance the current is suppressed. This is due to the effective electrochemical potential of the \qd~being ``kicked out'' of the transport window, see panel \panel{(c-2)}. Experimental observations of these features have been reported previously in Ref.~\cite{Wang2021}. Comparing with Fig.~\subfigref{fig:Figure2}{d}, we conclude that the arch-like resonance arise due to the intrinsic nonlinearities modeled by $\hat{H}_\text{duff}$ in Eq.~\eqref{eq:full_Hamiltonian}. The appearance of these arch-like resonance composes a main feature of our model. 

In the following section, we generalize our model to multiple electrochemical potentials of the \qd~entering the bias window and compare with experimental results. 

\section{Generalization to multiple QD electrochemical potentials \& experimental results}\label{Sec:Multilevel_gen}

\begin{figure*}[t]
    \centering
    \includegraphics[width=\textwidth]{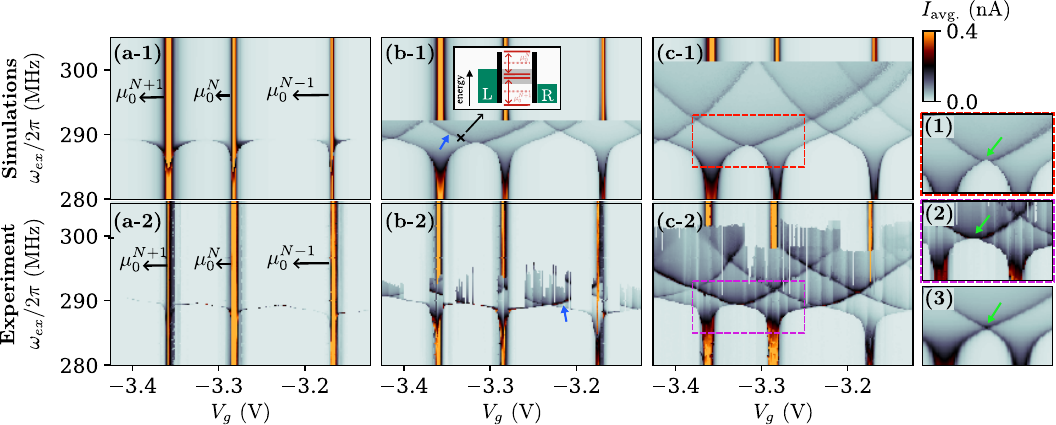}
    \caption{
    Top panels \textbf{(a-1)-(c-1)} showcase the simulated average current as a function of the gate voltage $V_g$ and the driving frequency $\omegaex$ (from low to high values) following the model described in Sec.~\ref{Sec:Multilevel_gen}. Left to right are for increasing external driving forces: 
    \textbf{(a-1)} $\F=0.06$ pN, \textbf{(b-1)} $\F=0.28$ pN, and \textbf{(c-1)} $\F=0.57$ pN. Here we consider three \qd~electrochemical potentials labeled as $\mu_0^{N+1}$, $\mu_0^{N}$, and $\mu_0^{N-1}$. The parameters used for these simulations are detailed in Table~\ref{tab:Simulation_parameters}.
    Bottom panels \textbf{(a-2)-(c-2)} are the measured average current as a function of the gate voltage $V_g$ and the driving frequency $\omegaex$ (from low to high values) with an applied bias of $0.25$ mV, the tunneling and mechanical extracted parameters are detailed in Appendix~\ref{Appendix:DataAnalysis}. Left to right describe an increase on driving power being \textbf{(a-2)} -76 dBm, \textbf{(a-2)} -66 dBm and \textbf{(a-3)} -53 dBm. 
    Panels \textbf{(1)}-\textbf{(3)} are a zoom-in of the current corresponding to the strong driving regime for the simulation considering the electromechanical back-action \textbf{(1)}, experimental \textbf{(2)} and simulation when no electromechanical back-action is considered \textbf{(3)}, the green arrows highlights the curvature of the $\I$ resonance corresponding to two consecutive electrochemical potential $\mu_0^{N+1}$ and $\mu_0^N$. All the panels share the same color-bar detailed in the upper right corner of the figure.}
    \label{fig:Figure4}
\end{figure*}

Up to now we have studied in detail the \NESS~response given by the model detailed in Sec.~\ref{Sec:Model} for different external driving forces. We identified the key features both in the displacement of the \resonator~(Sec.~\ref{Sec:NESS_response}) and the current for a single electrochemical potential (Sec.~\ref{Sec:el_current}). In this section we extend the model to take into account multiple electrochemical potentials in the \qd: $\{\mu^i_0\}_{i=N-.., N-1,N,N+1,N+2,...}$, see Fig.~\subfigref{fig:Figure1}{b}. As we will see later, this generalization is required to properly reproduced our experiments. If the displacement of the resonator (with respect to its zero-point motion) is small, the \qd~energy $\mu$ does not present strong deviations with respect to $\mu_0$. Thus it is safe to say that the electronic transport can only be due to a single electrochemical potential entering the transport window. On the contrary, as the driving force increases, so does the displacement, and therefore the change in $\mu$ can be large enough that different charged stated can contribute to the current.

To extend our model we first consider that each electrochemical potential $\mu^i_0$ has its own electromechanical coupling, such that Eq.~\eqref{eq:Chemical_eff} results in 
\begin{equation}\label{eq:Chemical_eff_multiple_levels}
    \mu^i(x)=\mu^i_0+ \hbar \omega \lambda^i x.
\end{equation}
Due to each electromechanical coupling, the displacement of the \resonator~is altered by the electronic transport through each of the multiple electrochemical potentials, thus the \EOM~given Eq.~\eqref{eq:displacement_EOM} is extended to
\begin{equation}\label{eq:displacement_EOM_multiple_levels}
    \ddot{x} + \delta \dot{x} + \omega^2 \Big[x+2\sum_i \lambda^i n^i(x) \Big] + \epsilon x^3 = f_{ex} \cos(\omegaex t),
\end{equation}
where $n^i(x)$ is the general version of Eq.~\eqref{eq:population_eq} as,
\begin{equation}\label{eq:population_eq_multiple_levels}
    n^i(x)= \frac{1}{2}+ \frac{\Gamma^i_R}{\pi \Gamma^i_{tot}}\Big[\arctan\left(\gamma^i_R(x)\right)+\frac{\Gamma^i_L}{\Gamma^i_R}\arctan\left(\gamma^i_L(x)\right)\Big].
\end{equation}
We note that we allow for each electrochemical potential $\mu^i_0$ to have its own tunnelings parameters, i.e. $\Gamma^i_{L/R}$, and thus $\gamma^i_\nu(x)=2(\mu_\nu-\mu^i(x))/(\hbar \Gamma^i_{tot})$. As in previous sections, in order to compute the \NESS~we simultaneously solve Eq.~\eqref{eq:Chemical_eff_multiple_levels}-\eqref{eq:population_eq_multiple_levels}.

Here, we directly study the current given that it is experimentally accessible and we later compare with measurements. In Appendix~\ref{Appendix:CNT_dis_Exps} a detailed analysis of the \resonator~displacement can be found. To account for all the contributions to transport arising from the different electrochemical potentials, Eq.~\eqref{eq:Average_current} is modified as
\begin{equation}\label{eq:Average_current_multiple_levels}
    \I= \lim_{\tau \rightarrow \infty} \int_\tau^{\tau+4\pi / \omega_0} \mathrm{d}t\, \sum_i I^i(x(t))
\end{equation}
with the instantaneous generalized current [from Eq.~\eqref{eq:current_eq}]
\begin{equation}\label{eq:current_eq_multiple_levels}
    I^i(x) =  \frac{e}{\pi} \frac{\Gamma^i_L\Gamma^i_R}{\Gamma^i_{tot}} \Big[\arctan\left(\gamma^i_L(x)\right)-\arctan\left(\gamma^i_R(x)\right)\Big].
\end{equation}

We compare the predictions of our generalized model with measurements of electron transport in a suspended \resonator~device. Fabrication, experimental details and data analysis can be found in Appendix~\ref{Appendix:Device_Fabrication_Details}. 
Varying the amplitude of the driving power $\P$, we explore different driving regimes of the electromechanical system.

Following a similar approach to that in Sec.~\ref{Sec:el_current}, we start by probing the device response at low driving power $\P$, which corresponds to a low driving force $\F$~\footnote{The conversion from the experimental driving power $\P$ to the driving force $\F$ considered in the model is non trivial, thus we estimate their values based on the observed features in the current.}. Fig.~\subfigref{fig:Figure4}{a-1} shows the simulated $\I$ (from Eq.~\eqref{eq:Average_current_multiple_levels}) as a function of gate voltage $V_g$ (which displaces the \qd~electrochemical potential $\{\mu_0^i\}_i$) and driving frequency $\omegaex$. Fig.~\subfigref{fig:Figure4}{a-2} shows the measured current. In both cases we observe Coulomb peaks corresponding to electrochemical potentials $\mu_0^{N+1}$, $\mu_0^{N}$, and $\mu_0^{N-1}$, and, as expected, the mechanical resonance exhibits softening at each of these peaks. We only showcase the $\I$ due to sweeping the $\omegaex$ from low to high values (as indicated in with red arrows), see Appendix~\ref{Appendix:bistable} for the results of the opposite sweeping direction.

Next, we increase the driving power $\P$ (and hence force $\F$) in Fig.~\subfigref{fig:Figure4}{b-1} (Fig.~\subfigref{fig:Figure4}{b-2}).
Here, for both the simulations and experiments, the softening of the mechanical resonance frequency is more evident and an arch-like structure begins to develop (highlighted with blue arrows). Each of the peaks in $\I$ can be traced back to the Coulomb peaks corresponding to $\mu_0^{N+1}$, $\mu_0^{N}$, and $\mu_0^{N-1}$. In between two peaks (marked with a black cross) an enhancement of $\I$ is observed. We attribute this feature to the fact that higher amplitudes of the  \resonator~motion are achieved with a higher driving $\P$($\F$), which in turn causes a larger shift in the electrochemical potentials of the \qd, allowing for both $\mu_0^{N+1}$ and $\mu_0^{N}$ to enter the transport window within one cycle of the \resonator~oscillation (see inset). In this way, the combination of these transport channels contribute to an increase of the net current. 

Finally, we further increase the driving power (force) to the strong driving regime, Fig.~\subfigref{fig:Figure4}{c-1} and~\subfigref{fig:Figure4}{c-2}. Here we see the arch-shaped structure fully develops. As detailed in the previous section, the observed arch-like response is due to the Duffing nonlinear contribution. Although here we are in the regime where the intrinsic \resonator~nonlinearities are predominant, we still note that the back action of the electronic transport plays a role. To see this, we zoom-in on a region corresponding to a crossing of electrochemical potentials marked with a black dashed square for the simulation \textbf{(1)} and experiment \textbf{(2)}. The enhancement of $\I$ corresponding to the contribution of two electrochemical potentials, namely $\mu_0^{N+1}$ and $\mu_0^N$, present a concave U-shape (highlighted with a green arrow), present both in our simulation and in the experiment. For comparison, in panel \textbf{(3)} we plot the case where we do not take into account the back action of the electronic transport into the displacement of the \resonator, as was done in previous works~\cite{Wang2021}. Here the concave shape is a V-shape (instead of a U-shape) with an enhancement of the $\I$ in the vertex, thus over estimating the electronic current.

\section{Conclusions}\label{Sec:Conclutions}

In this work, we studied the nonlinear response of an externally driven \resonator~mechanical resonator with a electrostatically defined \qd. The system displays the interplay of two distinct sources of mechanical nonlinearities, namely (i) coupling between the electronic transport and the motion and (ii) intrinsic mechanical contributions. We have presented a theoretical model that is able to capture these very different contributions by varying the external driving force. First, we considered the case where the electronic transport is given by a single electrochemical potential in the \qd. We found that, in the weak driving regime, the predominant nonlinearity is given by the electromechanical coupling, source (i). In such case, the resonance frequency presents the well-known softening dip given by the electromechanical coupling strength $\lambda$, consistent with previous works~\cite{Vigneau2022,Steele2009,Meerwaldt2012}. As the driving power is increased, we found that these dips persists although the mechanical nonlinearities, source (ii), start to dominate. A broadening of the resonance frequency was observed and it presented a frequency sweep hysteresis. In the strong driving regime, where source (ii) dominates, we have that the displacement response of the \resonator~is of a Duffing-type. Moreover, in the current through the device we observed the development of arch-like structures. This arch-like appearances constitutes the main feature of the presented model in the strong driving regime. We then extended our model to account for multiple electrochemical potentials of the \qd~inside of the bias window. We finally compare the prediction with device measurements of the current through a \resonator~and observed that they successfully qualitatively reproduce the experimental results and capture features that had previously been overlooked.

\begin{acknowledgments}
F.C. thanks Patrice Camati and L\'ea Bresque for useful discussions. S.S. acknowledges useful conversations with Karen Hovhannisyan, Felix Hartmann, and Joachim Ankerhold. This research was supported by grant number FQXi-IAF19-01 from the Foundational Questions Institute Fund, a donor-advised fund of the Silicon Valley Community Foundation. K.A. and N.A. acknowledge the support provided by funding from the Engineering and Physical Sciences Research Council IAA (Grant number EP/X525777/1). N.A. acknowledges the support from the Royal Society (URF-R1-191150), EPSRC Platform Grant (grant number EP/R029229/1) and the European Research Council (ERC) under the European Union’s Horizon 2020 research and innovation programme (grant agreement number 948932). N.A. and F.F. also acknowledge support from the European Union and UK Research \& Innovation (Quantum Flagship project ASPECTS, Grant Agreement No.~101080167). Views and opinions expressed are however those of the authors only and do not necessarily reflect those of the European Union, Research Executive Agency or UKRI. Neither the European Union nor UKRI can be held responsible for them. S.S. and J.A. acknowledge the support by Deutsche Forschungsgemeinschaft (DFG 384846402). J.M. acknowledges financial support from the Knut and Alice Wallenberg foundation through the fellowship program. GJM acknowledges support from the Australian Research Council Centre of Excellence for Engineered Quantum Systems (EQUS, CE170100009). J.T-B. acknowledges the support of the Alexander von Humboldt Foundation.

\end{acknowledgments}

\section{Author Contributions}

\textbf{S. Sevitz}: Model development (equal); Simulations (lead); Visualization (lead); Data Analysis (lead);  Writing – original draft (lead); Writing – review \& editing (equal). 
\textbf{K. Aggarwal}: Experimental data (lead); Visualization (supporting); Writing – original draft (supporting); Discussions (equal).
\textbf{J. Tabanera-Bravo}: Discussions (equal); Writing – review \& editing (equal).
\textbf{J. Monsel}: Model development (equal); Simulations (supporting); Discussions (equal); Writing – review \& editing (equal).
\textbf{F. Vigneau}: Fabrication (lead);Writing – review \& editing (equal).
\textbf{F. Fedele}: Discussions (equal); Writing – review \& editing (equal).
\textbf{J. Dunlop}: Discussions (equal); Writing – review \& editing (equal).
\textbf{J.M.R. Parrondo}: Discussions (equal).
\textbf{G.J. Milburn}: Model development (equal); Discussions (equal).
\textbf{J. Anders}: Supervision (equal); Discussions (equal); Writing – review \& editing (equal).
\textbf{N. Ares}: Supervision (equal); Discussions (equal).
\textbf{F. Cerisola}: Model development (equal); Simulations (supporting); Visualization (supporting); Data Analysis (supporting); Writing – review \& editing (equal).

\bibliography{references.bib}

\appendix
\onecolumngrid 

\section{Derivation of the CNT equation of motion}\label{Appendix:EOM_response}

In this Appendix we detail how to obtain Eq.~\eqref{eq:displacement_EOM} of the main text. To do so, we use Hamilton's equations for the full Hamiltonian detailed in Eq.~\eqref{eq:full_Hamiltonian}. This results in the following classical equation of motion
\begin{equation} 
    m \ddot{z}+ m \delta \dot{z} + m \omega^2 z+ \frac{\hbar \omega \lambda}{\zpm} n(z(t))  + \frac{\beta}{\zpm}  \left(\frac{z}{\zpm}\right)^3 = \F \cos(\omegaex t).
\end{equation}
Note that we have included a friction term $m\delta\dot{z}$ to account for the intrinsic sources of dissipation of the mechanical oscillator. Then, defining the normalized displacement with respect to the zero point motion, $x=z/z_\text{ZPM}$, one arrives to Eq.~\eqref{eq:displacement_EOM} of the main text.

\section{Numerical calculation of the non-equilibrium steady state}\label{Appendix:Num_details}

\begin{figure}[ht]
    \centering
    \includegraphics[width=.9\linewidth]{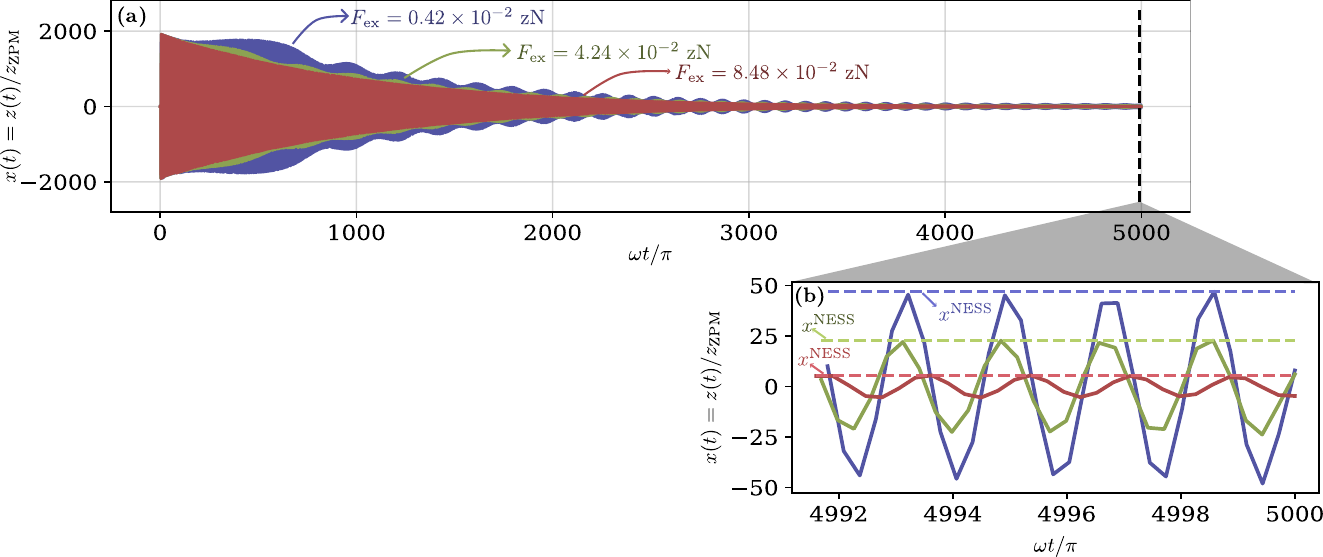}
    \caption{
    \textbf{(a)} Time dynamics for the displacement of the \resonator~with respect to its zero -point motion $\zpm$ for three external driving forces: $\F=0.4$ fN (purple), $\F=4$ fN (green) and $\F=8$ fN (red). In these panels the driving frequency $\omegaex$ was set to $1$ Hz and the \qd~energy to $\mu_0=$ meV.
    \textbf{(b)} Zoom-in in to large time scale highlighted with a black dashed line in panel \textbf{(a)}. With horizontal dashed lines we characterize the \NESS~displacement of the \resonator~for each driving force. 
    Parameters: $\zpm=1$ pm , $\omega=100$ MHz, 
    $m=0.5$ fg, $\delta=70 $ kHz, $\beta=4.69 \times 10^{-5}$ qJ, $\gw=$ 1.73, $\mu_L=-\mu_R= 0.1$ meV and $\Gamma_L/2\pi=\Gamma_R/2\pi= 30$ GHz.} 
    \label{fig:Figure5_appendix}
\end{figure}

In order to obtain the \NESS~of the semiclassical system, we simultaneously solve Eqs.~\eqref{eq:displacement_EOM}-\eqref{eq:population_eq} of the main text. This was carried out using the ``DifferentialEquations.jl'' Julia package version 7.8.0~\cite{Rackauckas2017}. The solver used is ``Verner's `Most Efficient' 9/8 Runge-Kutta method'' with initial time step $1\times10^{-10}$, absolute tolerance $1\times 10^{-5}$, and relative tolerance $1\times 10^{-5}$. We integrated the equations of motion, obtaining the full dynamics of the \resonator~for long times until the \NESS~was reached, i.e. about $5000\pi/\omega$ for the parameters used. In Fig.~\subfigref{fig:Figure5_appendix}{a} we plot an example of the dynamics for the parameters of Fig.~\ref{fig:Figure2} of the main text. Here, we choose $\omegaex/\omega=1$ and $\mu_0=0$ meV. As the \resonator~is externally driven, we expect the \NESS~to be of the form $x(t\rightarrow \infty)=x^\text{NESS}\cos(\omegaex t)$. The amplitude $x^\text{NESS}$ is then obtained from the amplitude of long time dynamics, as shown in Fig.~\subfigref{fig:Figure5_appendix}{b}.

\section{Device fabrication \& experimental details}\label{Appendix:Device_Fabrication_Details}

The device is a nanoelectromechanical resonator formed by suspending a carbon nanotube on metal electrodes (Fig.~\subfigref{fig:Figure6_appendix}{a}). A bias voltage $V_\text{SD}=\Delta\mu/e$ is applied from source to drain electrodes, driving a current $I_{DC}$ through the nanotube. The carbon nanotube is suspended above five gate electrodes with voltages $V_\text{G1-G5}$. The mechanical oscillations of the nanotube are excited by a radiofrequency (rf) drive, with frequency $f_\text{ex}$ and driving power $\P$, connected to one of the gates. The gates are further used to define a \qd~along the carbon nanotube. All experiments in this work are carried out in a dilution refrigerator with a base temperature of 60 mK.

The suspended carbon nanotube device is the same as the one presented in Ref.~\cite{Vigneau2022} but measured in a different cool down. The device is fabricated on a high resistance Si/SiO$_\text{2}$ substrate by patterning Au/Cr electrodes with electron beam lithography. Carbon nanotubes are grown by chemical vapor deposition on a separate quartz substrate using nanoparticles of Al$_\text{2}$O$_\text{3}$, Fe(NO$_\text{3}$) and MoO$_\text{2}$(acac)2 as catalyst and mechanically transferred to the Si chip. 

\section{Characterization of the device}\label{Appendix:DataAnalysis}

\begin{figure}[ht]
    \centering
    \includegraphics[width=\linewidth]{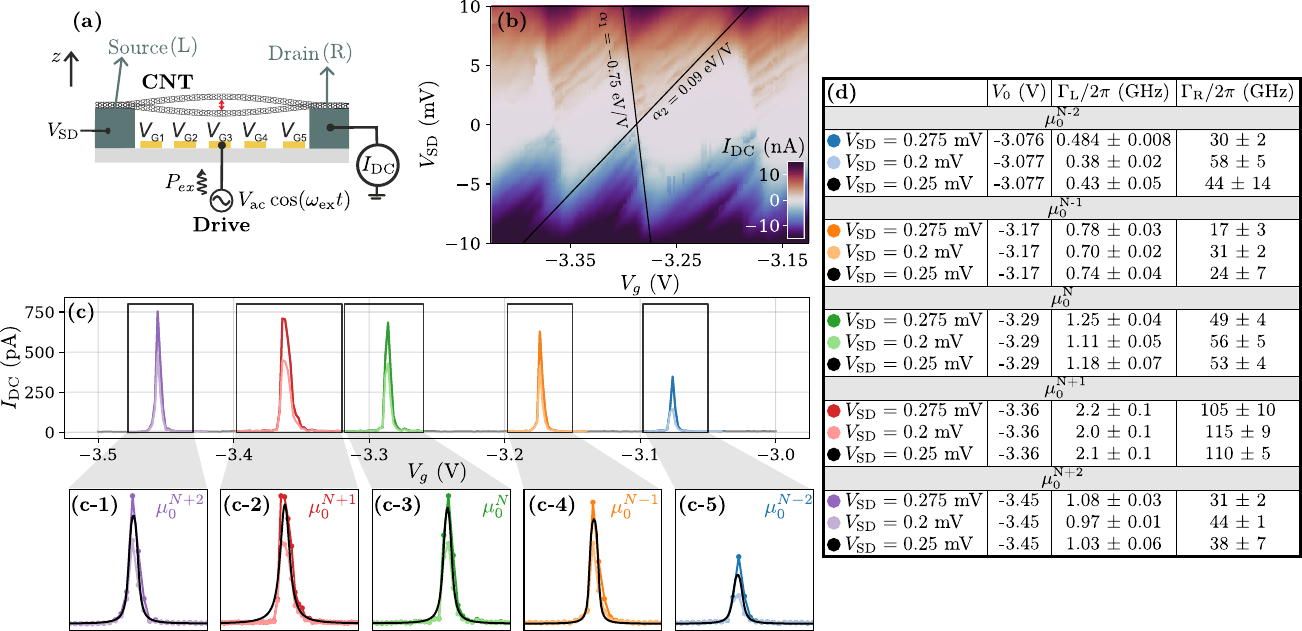}
    \caption{
    \textbf{(a)} Schematic of the experimental device. A carbon nanotube is suspended between the source and drain metal electrodes. Five gate electrodes with voltages $V_{\text{G1-G5}}$ are used to define a quantum dot. A bias voltage $V_\text{SD}=\Delta\mu/e$ drives a current $I_{DC}$ through the carbon nanotube. An rf drive $V_{ac} \cos{\omegaex t}$ with power $\P$ excites the mechanical motion of the carbon nanotube.
    \textbf{(b)} Measured $I_\text{DC}$ as a function of $V_\text{SD}$ and $V_g=V_{G1}$ showing Coulomb diamonds. The slopes $\alpha_1$ and $\alpha_2$ of the Coulomb diamonds, highlighted in black, are used to estimate the lever arm $\alpha$.
    \textbf{(c)} Coulomb peaks for the range of $V_g=V_\text{G1}$ under consideration in the main text. The darker color trace corresponds to $V_\text{SD}= 0.275$ mV and the lighter color trace corresponds to $V_\text{SD}= 0.2$ mV. The bottom panel shows a zoom in of the Coulomb peaks. The solid black lines fits evaluated using Eq.~\eqref{eq:current_eq} (and setting $\gw=0$) with parameters exhibited in panel \textbf{(d)}.
    }
    \label{fig:Figure6_appendix}
\end{figure}

Now we detail how we extracted the tunneling and coupling parameters from the experimental data to later employ in the model and numerical calculations. 

\medskip

We first extract the lever arm $\alpha = eC_\text{G}/C$, where $C_\text{G}$ is the capacitance between the gate and the \qd, and $C = C_\text{G} + C_\text{S} + C_\text{D}$ is the total capacitance between the gate, source and drain. Similarly to~\cite{Vigneau2022}, we calculate the two slopes $\alpha_1 = -eC_\text{G}/(C-C_\text{S})$ and $\alpha_2 = eC_\text{G}/C_\text{S}$ of the Coulomb diamonds (without an external driving) as shown in Fig.~\subfigref{fig:Figure6_appendix}{b}. The lever arm $\alpha$ is determined as~\cite{Hanson2007}
\begin{equation}
    \alpha = \frac{\alpha_1\alpha_2}{\alpha_1-\alpha_2} = 0.083 \text{ eV/V}.
\end{equation}

\medskip

Next, we extract the tunneling parameters for the \qd~electrochemical potentials $\mu_0^{N+2}$--$\mu_0^{N-2}$. To estimate the tunneling parameters for the bias voltage $ V_\text{SD} = 0.25$ mV used in Sec.~\ref{Sec:Multilevel_gen}, we average the value by fitting Coulomb peaks to the nearby valued of $V_\text{SD} = 0.2$ mV and $V_\text{SD} = 0.275$ mV. The measured Coulomb peaks are presented in Fig.~\subfigref{fig:Figure6_appendix}{c} (with external driving $\P=0$) from which the tunneling rates are obtained by fitting each peak to Eq.~\eqref{eq:current_eq} and setting $\gw=0$.
In Fig.~\subfigref{fig:Figure6_appendix}{d} we show the results of the fits.

\medskip

\begin{figure}[ht]
   \centering
   \includegraphics[width=\linewidth]{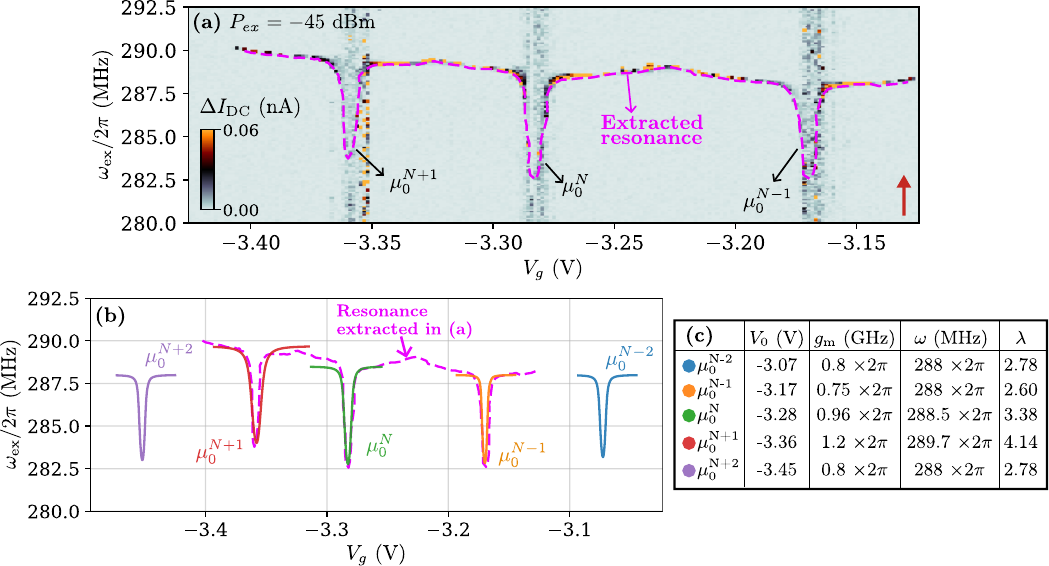}
   \caption{
   \textbf{(a)} Differential current response to the lowest driving power, i.e. $-45$ dBm, as a function of gate voltage $V_g$ and driving frequency $\omegaex$ swept from low to high values as indicated with a red arrow. The extracted resonance for the $\mu_0^{N+1}-\mu_0^{N-1}$ \qd~electrochemical potentials is highlighted with a dashed magenta line.
   \textbf{(b)} Softening of the mechanical resonance for different Coulomb peaks extracted from panel \textbf{(a)} (shown in dashed magenta). The solid lines show the calculated softening of the mechanical resonance using parameters in panel \textbf{(c)} following Ref.~\cite{Vigneau2022}.
   \textbf{(c)} Parameters characterizing the coupling of mechanical motion to the electron transport obtained from panel \textbf{(a)}. The parameters for peaks corresponding to $\mu_0^\text{N-2}$ and $\mu_0^\text{N+2}$ are estimated based on experimentally realistic values.
    }
    \label{fig:Figure7_appendix}
\end{figure}

To compute the electromechanical coupling strength $g_\text{m}=\omega \gw$ we turn on the external drive with power $\P=-45$ dBm and follow the same procedure as in Ref.~\cite{Vigneau2022}. Due to the multiple electrochemical potentials of the \qd~participating in the transport, we determine individually each of the electromechanical coupling strengths $g_\text{m}^i$, corresponding to the $i$-th electrochemical potential. In Fig.~\subfigref{fig:Figure7_appendix}{a} we showcase the frequency response of the \resonator~ as a function of the gate voltage $V_g$ and external driving frequency $\omegaex$. From here we can extract the softening of the mechanical resonance of the peaks $\mu_0^{N+1}$--$\mu_0^{N-1}$. We find the values of $\{g_\text{m}^i\}_{i=N+1,N,N-1}$ by qualitatively matching the experimental data as shown in Fig.~\subfigref{fig:Figure7_appendix}{b}. The extracted parameters can be found in ~\subfigref{fig:Figure7_appendix}{c}. The parameters for peaks corresponding to $\mu_0^{N+2}$ and $\mu_0^{N-2}$ were estimated based on experimentally realistic values.

\medskip

\begin{figure}[ht]
    \centering
    \includegraphics[width=.8\linewidth]{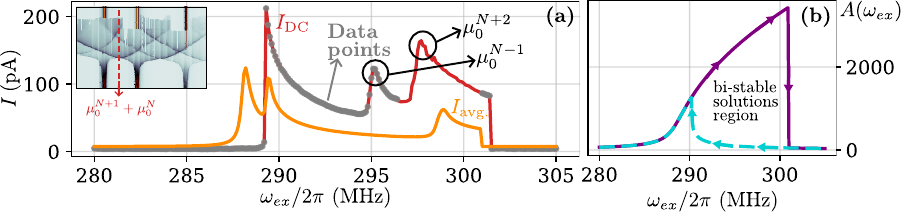}
    \caption{
    \textbf{(a)} $I_\text{DC}$ as a function of driving frequency $\omegaex$. The red trace is the measured mechanical resonance for driving power $\P=-53$ dBm and gate voltage $V_{G1}=-3.34$V, when the Duffing nonlinearity is prominent. The yellow trace is the calculated $\I$ calculated using the guess values ($\epsilon=4.92\times10^9$ Hz$^2$ and $\delta=2.09$ MHz) and normalized force $f_{ex}=2 \times 10^{18}$ Hz$^2$.
    \textbf{(b)} Displacement of the \resonator~with the simplification made in Eq.~\ref{eq:aprox_dispaclemt} used to compute $\I$ of panel \textbf{(a)}.
    }
    \label{fig:Figure8_appendix}
\end{figure}

The final step is to characterize the Duffing parameter $\epsilon$ and damping $\delta$ of the \resonator~as defined in Eq.~\eqref{eq:displacement_EOM}. In principle, these parameters should be obtained by fitting the displacement of the \resonator. However, in the experiment we only have access to the current $I_\text{DC}$ through the nanotube. We therefore estimate these parameters so as to be able to reproduce the current intensity and broadening observed in the experiment 
at a fixed gate voltage $V_g$. Fig.~\subfigref{fig:Figure8_appendix}{a} shows the current $I_\text{DC}$ for driving power $\P = -53$ dBm and $V_g = -3.34$ V. We see that the current $I_\text{DC}$ presents three peaks, corresponding to the \qd~electrochemical potentials $\mu_0^{N+1}+\mu_0^{N}$, $\mu_0^{N-1}$, and $\mu_0^{N+2}$. The various parameters characterizing the electron transport through the \qd~and the mechanical motion of the \resonator~are intricately linked to each other. Therefore, finding each value of these parameters for all the combinations of the other becomes an onerous challenge. Thus, we make a few simplifications to estimate the resonator parameters:
\begin{enumerate}
    \item In the experiment, the frequency of the mechanical resonance of the nanotube changes as a function of the gate voltage $V_g$, see Fig.~\subfigref{fig:Figure7_appendix}{a}. The first simplification that we consider is that the mechanical resonance frequency $\omega$ of the \resonator~is independent of $V_g$, such that for all peaks $\omega/2\pi = 289$ MHz.
    Experimentally, this can be seen in Fig.~\ref{fig:Figure4} as the overall linear trend of the frequency to decrease as $V_g$ increases. Indeed, this is why in the simulations after each dip the frequency goes back up to the same value as before,
    while in the experiment it goes to a value slightly smaller. The dependence of the frequency on $V_g$ has been extensively studied previously elsewhere, see e.g. Refs.~\cite{Meerwaldt2012,Meerwaldt2012Chapter}. Here, we choose to exclude it from our model to simplify the number of free parameters, since except for this overall trend, it does not affect the shape of the frequency dips nor the broadening effects that are the main focus of this work. Adding this effect to our model could be done straightforwardly.
    \item Since Eqs.~\eqref{eq:displacement_EOM_multiple_levels}--\eqref{eq:current_eq_multiple_levels} do not allow for a  simple analytical expression for the current in the steady-state, we consider that in the \NESS~the amplitude of the \resonator~follows the response of a Duffing oscillator, that is
    \begin{align}\label{eq:aprox_dispaclemt}
        z(t) =A(\omega_\text{ex}) \cos (\omega_\text{ex} t) \quad \text{with} \quad
        A(\omega_\text{ex}) =\frac{\omega_\text{ex}}{2\pi\sqrt{\left(\omega_\text{ex}^2-\omega_\text{m}^2-\frac{3}{4}\epsilon A(\omega_\text{ex})^2\right)^2+\left(\delta \omega_\text{ex}\right)^2}}.
    \end{align}
    Then, given $\epsilon$ and $\delta$, the \NESS~current is given by replacing this value of $z$
    in Eq.~\eqref{eq:Chemical_eff_multiple_levels} and Eq.~\eqref{eq:current_eq_multiple_levels}.
\end{enumerate}
Under these simplifications, for every value of $\epsilon$ and $\delta$ we produce an approximate current response and compare with the experimental measurements. In Fig.~\subfigref{fig:Figure8_appendix}{a} in orange we plot the result for the best set of values, $\epsilon=4.92\times10^9$ Hz$^2$ and $\delta=2.09$ MHz. The corresponding displacement of the \resonator~using Eq.~\eqref{eq:aprox_dispaclemt} is shown in Fig.~\subfigref{fig:Figure8_appendix}{b}. We find that with these simplifications, the model is qualitatively able to capture the role of different nonlinearities in the experimental data. 
This allows us to understand a hybrid nano-electromechanical system with a simple theoretical model. In Table~\ref{tab:Simulation_parameters} we summarized all the parameters used in order to compute the Fig.~\subfigref{fig:Figure4}{a-1}-
Fig.~\subfigref{fig:Figure4}{a-3}.

\medskip

It is important to note that in our experimental data set, we observe both softening and hardening of the mechanical resonance. This is not captured in the model because the Duffing parameters are fixed to one of these cases. We attribute this change of value and sign of the non-linearity to different gate voltage configurations and driving power, as this phenomenon has been previously observed and characterized both in CNT and other nanomechanical systems~\cite{Poot2012,Samanta2018,Steeneken2021,Steele2009}.

\begin{table}[ht]
    \begin{tabular}{ cc }  
    \begin{tabular}{|l|c|c|c|c|c|}
        \hline
        $V_\text{SD}=0.25$ mV &$V_0$ (V) & $\Gamma_\text{L}/ 2\pi$ (GHz) & $\Gamma_\text{R}/ 2\pi$ (GHz) & $\gw$& $\omega $ (MHz)  \\ \hline
         $\mu_0^{N-2}$& -3.07135 & 0.430264 & 44.2383 & 2.78 & 289 $\times 2\pi$\\
         $\mu_0^{N-1}$&-3.16863&0.737844 & 24.0938 & 2.60 & 289 $\times 2\pi$\\
         $\mu_0^{N}$& -3.2814 & 1.17658 & 52.808 & 3.38& 289 $\times 2\pi$\\
         $\mu_0^{N+1}$&-3.35679 & 2.13673 & 110.354 & 4.14& 289 $\times 2\pi$\\
         $\mu_0^{N+2}$& -3.45068 & 1.02658& 37.7511 & 2.78& 289 $\times 2\pi$\\ 
         \hline
    \end{tabular} &  
    \begin{tabular}{ |c|c| } 
        \hline
        $\epsilon$ & $4.92\times10^9$ Hz$^2$ \\
        $\delta$ & $2.09$ MHz \\
         $\zpm$ & $(0.68 \pm 0.04)$ pm \\
         $m$ & 63 ag \\
         $\alpha$ & 0.083 eV/V \\
        \hline
        \end{tabular} \\
    \end{tabular}
    \caption{Summery of parameters extracted form the device used to compute the simulations it Fig.~\subfigref{fig:Figure4}{a-1}-Fig.~\subfigref{fig:Figure4}{c-1}. We take the $\zpm$ value from Ref.~\cite{Vigneau2022}, and use the $\omega$ to determine the \resonator~mass.}
    \label{tab:Simulation_parameters}
\end{table}

\section{CNT displacement for the estimated device parameters}\label{Appendix:CNT_dis_Exps}

\begin{figure}[ht]
    \centering
    \includegraphics[width=0.7\linewidth]{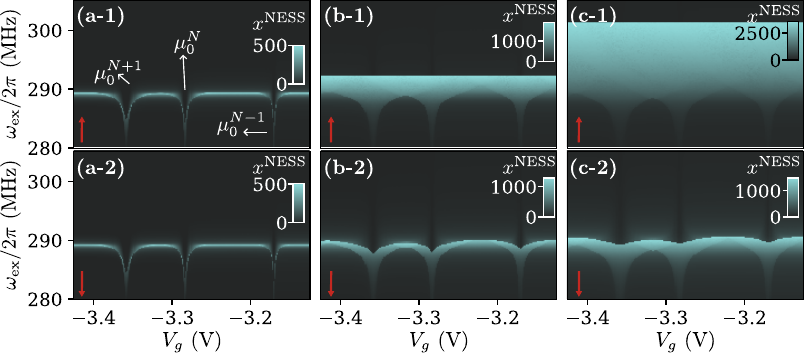}
    \caption{
    Displacement of the \resonator~with respect to its zero point motion $x^{NESS}=z^{NESS}/\zpm$ in the \NESS~as a function of both the external driving frequency $\omegaex$ and gate voltage for increasing driving forces as:  \textbf{(a-1)/(a-2)} $\F=0.06$ pN, \textbf{(b-1)/(b-2)} $\F=0.28$ pN and \textbf{(c-1)/(c-2)} $\F= 0.57$ pN.
    Top panels correspond to the external driving frequency $\omegaex$ driving from low to high powers (per indicated with a red arrow) and bottom panels correspond to the $\omegaex$ swept in the opposite direction. 
    }
    \label{fig:Figure9_appendix}
\end{figure}

In Fig.~\ref{fig:Figure9_appendix} we plot the \resonator~displacement for the parameters estimated from the experimental measurements shown in
Fig.~\ref{fig:Figure4}
as a function of the gate voltage $V_g$ and the driving frequency $\omegaex$ and for increasing external driving force.
The interpretation of the observed behavior is analogous to what was discussed in Sec.~\ref{Sec:NESS_response} for the single electrochemical potential case. Here, we moreover incorporate the effect of having multiple \qd~electrochemical potentials within the bias window, including the effect of the back action on the \resonator~displacement from each possible transition, as was covered in detail in Sec.~\ref{Sec:Multilevel_gen}.

\section{Bistable response and frequency sweeping hysteresis}\label{Appendix:bistable}

\begin{figure}[ht]
    \centering
    \includegraphics[width=.8\linewidth]{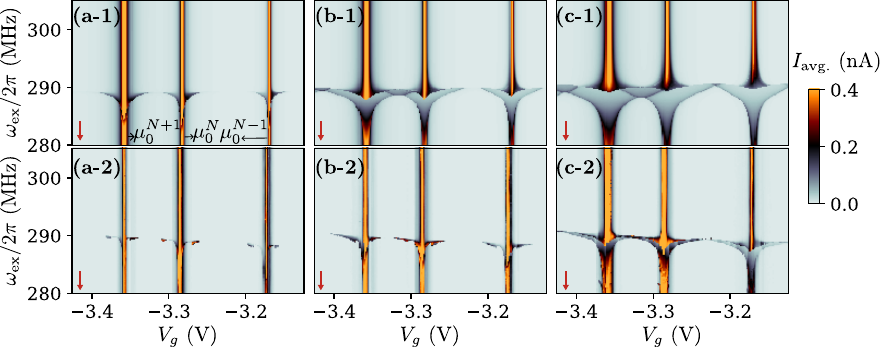}
    \caption{Similar to Fig.~\ref{fig:Figure4}, however, the external frequency $\omegaex$ is swept from high to low values. The top panels \textbf{(a-1)-(c-1)} shows the calculated $\I$ for external driving forces $\F=0.06$ pN, $\F=0.28$ pN and $\F= 0.57$ pN respectively. While the bottom panels \textbf{(a-2)-(c-2)} showcase the measured current $I_\text{DC}$ for different driving powers $\P = -76$ dBm, $\P= -66$ dBm, and $\P = -53$ dBm respectively. }
    \label{fig:Figure10_appendix}
\end{figure}

In Fig.~\ref{fig:Figure4} of the main text, we showed the experimentally measured current and the model predictions when the frequency $\omegaex$ of the external drive is swept from low to high frequency. In Fig.~\ref{fig:Figure10_appendix} we now plot instead the calculated (top panels) and measured (bottom panels) $\I$ response when the external driving frequency $\omegaex$ is swept from high to low frequency. Due to the strong nonlinearity of the system, a hysteresis in the response is observed. The dependence on sweeping direction of the measured $\I$ is weaker for the lowest driving force (power) in Fig.~\subfigref{fig:Figure10_appendix}{a-1}(Fig.~\subfigref{fig:Figure10_appendix}{a-2}) and becomes noticeably significant as the driving force (power) increases in Fig.~\subfigref{fig:Figure10_appendix}{b-1} and~\subfigref{fig:Figure10_appendix}{c-1}(Fig.~\subfigref{fig:Figure10_appendix}{b-2} and~\subfigref{fig:Figure10_appendix}{c-2}).

\end{document}